\documentclass{article}
\usepackage{amsmath}
\usepackage{amssymb}
\usepackage{graphicx}
\usepackage{hyperref}
\usepackage{geometry}
\usepackage{color}
\usepackage{tabularray}
\usepackage{array}
\usepackage{bigstrut}
\usepackage{multirow}
\usepackage{bm}
\usepackage{booktabs}
\usepackage[numbers]{natbib}

\begin{document}

\title{Enhancing Octree-based Context Models for Point Cloud Geometry Compression with Attention-based Child Node Number Prediction}
\author{Chang Sun, Hui Yuan, Xiaolong Mao, Xin Lu, and Raouf Hamzaoui}
\date{}
\maketitle

\begin{abstract}
 In point cloud geometry compression, most octree-based context models use the cross-entropy between the one-hot encoding of node occupancy and the probability distribution predicted by the context model as the loss. This approach converts the problem of predicting the number (a regression problem) and the position (a classification problem) of occupied child nodes into a 255-dimensional classification problem. As a result, it fails to accurately measure the difference between the one-hot encoding and the predicted probability distribution. We first analyze why the cross-entropy loss function fails to accurately measure the difference between the one-hot encoding and the predicted probability distribution. Then, we propose an attention-based child node number prediction (ACNP) module to enhance the context models. The proposed module can predict the number of occupied child nodes and map it into an 8-dimensional vector to assist the context model in predicting the probability distribution of the occupancy of the current node for efficient entropy coding. Experimental results demonstrate that the proposed module enhances the coding efficiency of octree-based context models.
\end{abstract}

\section{Introduction}
Owing to their flexibility in representing 3D objects and scenes, 3D point clouds have been widely used in immersive communication \cite{1}, virtual reality (VR) \cite{2}, automated driving \cite{3}, etc. Since point clouds entail huge data volumes, efficient compression algorithms \cite{4}\cite{5}\cite{6} are essential to reduce their storage and transmission costs. However, this presents a significant challenge due to their sparsity and irregular structure.

One of the most effective point cloud compression techniques is the geometry-based point cloud compression (G-PCC) standard \cite{7} developed by the moving picture experts group (MPEG). G-PCC uses an octree encoding method \cite{8} that partitions an input point cloud into nodes and relies on hand-crafted context models \cite{9}\cite{10}\cite{11}\cite{12} for entropy coding, enabling lossless compression of the point cloud. The performance of the context model significantly affects the coding efficiency. 

Recently, deep learning \cite{13}\cite{14}\cite{15}\cite{16}\cite{17} has been widely applied to point cloud compression. Some of the most successful methods learn context models for octree-based coding. For example, Huang \textit{et al.} \cite{18} proposed a context model called OctSqueeze. Although OctSqueeze is based on a simple multi-layer perceptron (MLP) and only considers ancestor nodes as context, it shows better performance compared to hand-crafted context models used in G-PCC. Biswas \textit{et al.} \cite{19} proposed a multi-sweep deep entropy model (MuSCLE) that constructs a context based on inter-frame correlation. However, this work overlooks the sibling nodes in the current frame. Que \textit{et al.} \cite{20} proposed a two-stage deep learning framework called VoxelContext-Net, which introduces $9\times9\times9$ sibling nodes as context. However, as the nodes become finer, the receptive field of the context becomes smaller, deteriorating the efficiency of entropy coding. Fu \textit{et al.} \cite{21} proposed an attention-based context model called OctAttention, which introduces thousands of sibling nodes and their ancestor nodes as context. However, OctAttention decodes the point cloud node by node, leading to high decoding complexity. Song \textit{et al.} \cite{22} proposed an  efficient hierarchical entropy model (EHEM) that introduces a hierarchical attention and a grouped context structure. Compared with OctAttention, EHEM maintains the global context receptive field while significantly enhancing the speed of encoding and decoding. 

The context models \cite{18}\cite{19}\cite{20}\cite{21}\cite{22} focus on proposing a better network structure with more context information. However, as loss function, they use a cross-entropy, which does not accurately measure the difference between the label and the predicted probability distribution. Specifically, the occupancy of a node is transformed into a 255-dimensional one-hot encoding, which serves as the training label. Then, the cross-entropy between this label and the probability distribution estimated by the context model is used as the loss. This loss, which is typically used in classification problems, can measure the positional differences between child nodes. However, it is not suitable to measure the difference between the predicted number and the actual number of occupied child nodes, which constitutes a regression problem. In \cite{23}, we proposed to add an MLP branch that predicts the occupancy of child nodes and use it as a feature to assist training. However, the predicted occupancy of child nodes does not directly correspond to the number of occupied child nodes.

In this letter, we propose an attention-based module aimed at enhancing learning-based context models in octree-based geometry compression by directly predicting the number of occupied nodes. The proposed module is general and can enhance the performance of a wide range of context models. The main contributions can be summarized as follows.

\begin{enumerate}

\item {We analyze why the cross-entropy loss based on one-hot encoding fails to measure the difference between the label and the probability distribution predicted by the context models.}
\item {We introduce an attention-based child node number prediction (ACNP) module to predict the number of occupied child nodes and map it into an 8-dimensional vector containing the information about the number of occupied child nodes. The resulting 8-dimensional vector then serves as a feature to assist in training the context model.}
\item {We demonstrate the effectiveness of the ACNP module by applying it to octree-based context models, specifically OctAttention and OctSqueeze. Experimental results on the MPEG 8i, MVUB and SemanticKITTI datasets show that adding the ACNP module can improve coding efficiency in lossless geometry compression.}
\end{enumerate}

The remainder of this letter is organized as follows. In Section \uppercase\expandafter{\romannumeral2}, we first analyze why the cross-entropy loss cannot accurately measure the difference between the label and the predicted probability distribution. Then, we describe the proposed ACNP module in detail. Experimental results and conclusions are given in Section \uppercase\expandafter{\romannumeral3} and Section \uppercase\expandafter{\romannumeral4}, respectively.

\section{PROPOSED METHOD}
\subsection{Problem Analysis}
In octree-based geometry compression, the features of encoded nodes are used as context to predict the probability distribution for entropy coding. The probability distribution predicted by the context model for the current node, $\boldsymbol{X_i}$, can be expressed as
\begin{equation}
\boldsymbol{P_i}=G(\boldsymbol{c_i};\boldsymbol{\omega}),
\end{equation}
where $G$ is the context model, $\boldsymbol{c_i}$ is the context of node $\boldsymbol{X_i}$, and $\boldsymbol{\omega}$ denotes the set of parameters of the context model that need to be learned. The 8 child nodes of $\boldsymbol{X_i}$ give 256 possible occupancy configurations ($x_{1},x_{2},...,x_7,x_8$), where $x_r \in \{0,1\}, r=1,...,8$, with 0 meaning empty node and 1 meaning occupied node. Thus, excluding the configuration where all child nodes are empty, $\boldsymbol{P_i}$ consists of the 255 probabilities $p_i(j),j=1,2...,255$, where $j$ is the decimal representation of the occupancy configuration ($x_{1},x_{2},...,x_7,x_8$) and $p_i(j)$ is the predicted probability of the occupancy configuration corresponding to $j$. The cross-entropy loss between the label and $\boldsymbol{P_i}$ can be expressed as
\begin{equation}
Loss_{ce} = -\sum_{i=1}^{m}\sum_{j=1}^{255}  e_i(j)\log_2p_i(j),
\end{equation}
where $m$ is the total number of nodes in the octree, and $e_i(j)$ represents the value of the label at integer $j$. As the label is based on one-hot encoding, $e_i(j)$ equals 1 if the actual occupancy configuration corresponds to $j$ and 0, otherwise. Thus, the cross-entropy loss simplifies to 
\begin{equation}
Loss_{ce} = -\sum_{i=1}^{m}\log_2{p_i},
\end{equation}
where $p_i$ is the predicted probability of the actual occupancy configuration of node $\boldsymbol{X_i}$. For example, as shown in Fig. 1, the cross-entropy loss between $\boldsymbol{A}$ and the label, and that between $\boldsymbol{B}$ and the label are identical according to (3). However, $\boldsymbol{A}$ assigns a higher probability to the occupancy configurations, i.e. the \textit{1st} and \textit{3rd} ones, which are closer to the actual occupancy \textit{00000010}. Specifically, $\boldsymbol{A}$ assigns a higher probability to the \textit{1st} occupancy, \textit{00000001}, where the number of occupied child nodes matches the actual occupancy, differing only in the positions of the occupied child nodes. On the other hand, $\boldsymbol{B}$ assigns a higher probability to the \textit{253th} occupancy configuration, \textit{11111101}, where both the number and position of occupied child nodes are different. As $\boldsymbol{A}$ assigns a higher probability to the occupancy configurations that are closer to the actual occupancy compared to $\boldsymbol{B}$, $\boldsymbol{A}$ is closer to the label, contradicting the computed cross-entropy loss. The above analysis demonstrates that cross-entropy loss cannot accurately measure the difference between the label and the predicted probability distribution, which is not conducive to the learning of context models.

\begin{figure}[!t]
\centering
\includegraphics[width=8cm]{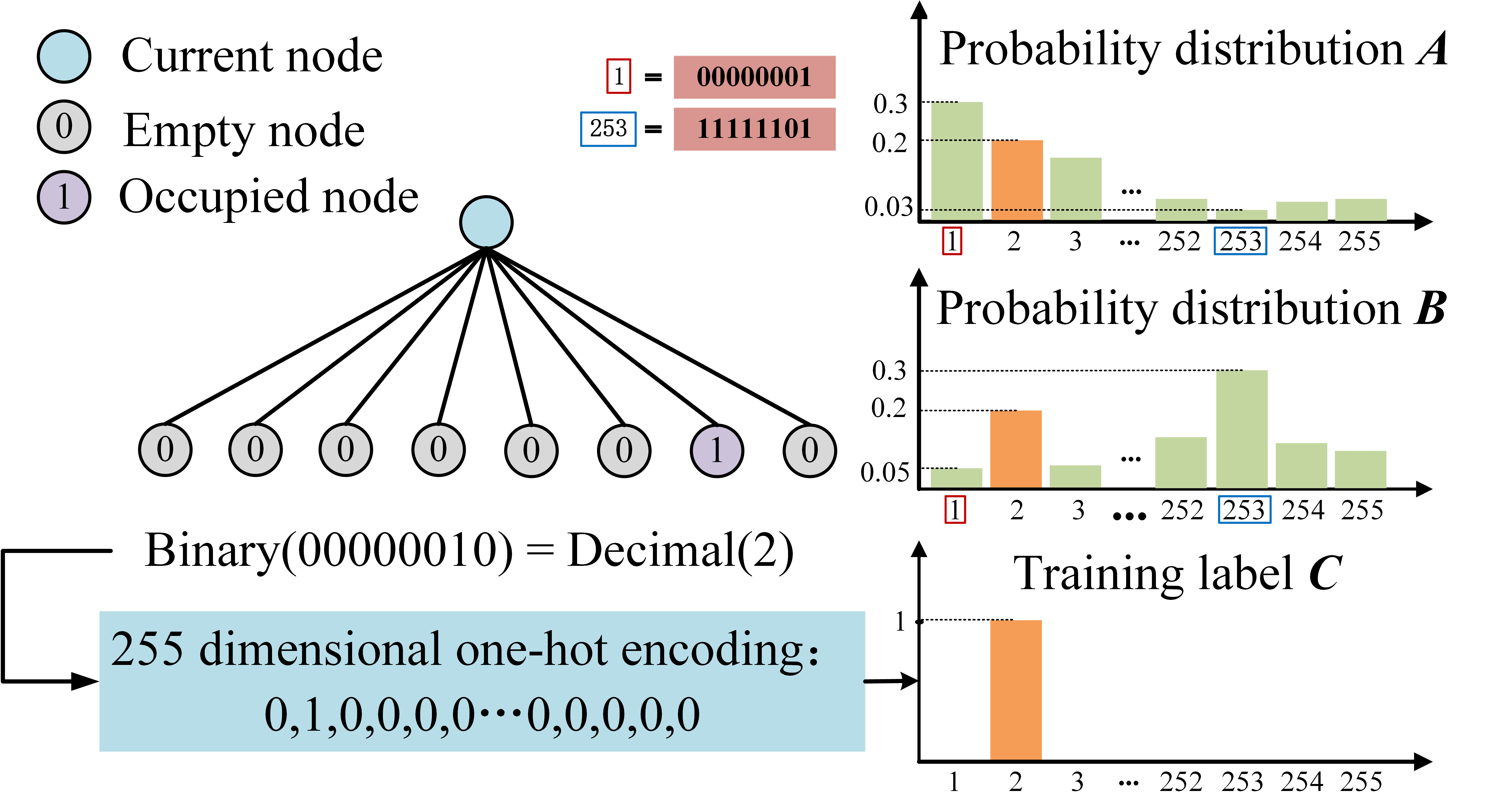}
\vspace{-0.3cm}%%减小图片上间隔
\caption{Problem of cross-entropy loss. The occupancy of the child nodes can be represented in binary as \textit{00000010}, which is equivalent to the decimal value 2. Therefore, the training label is a one-hot encoding with a value of 1 in the second dimension and 0 in all other dimensions. $\boldsymbol{A}$ and $\boldsymbol{B}$ are two probability distributions predicted by some context models. The cross-entropy loss between $\boldsymbol{A}$ and the label and that between $\boldsymbol{B}$ and the label are identical.}
\vspace{-0.3cm}
\label{fig1}
\end{figure}

The reason for the aforementioned issue is that the cross-entropy loss is only effective when the number of occupied child nodes is accurately predicted. Specifically, the occupancy of the current node contains the positions and number of the occupied child nodes, but the cross-entropy loss transforms the predicted probability distribution of occupied child nodes into a 255-dimensional classification problem according to (3). Due to the proximity of child nodes in 3D space, determining the position of occupied child nodes becomes a classification problem when the number of occupied child nodes is known. In this case, cross-entropy loss is applicable. However, the number of occupied child nodes is unknown, and it is a regression problem that cross-entropy loss cannot measure according to (3). As a result, cross-entropy loss fails to accurately measure the difference between the label and the predicted probability distribution. 

 \begin{figure}[!t]
\centering
\includegraphics[width=8cm]{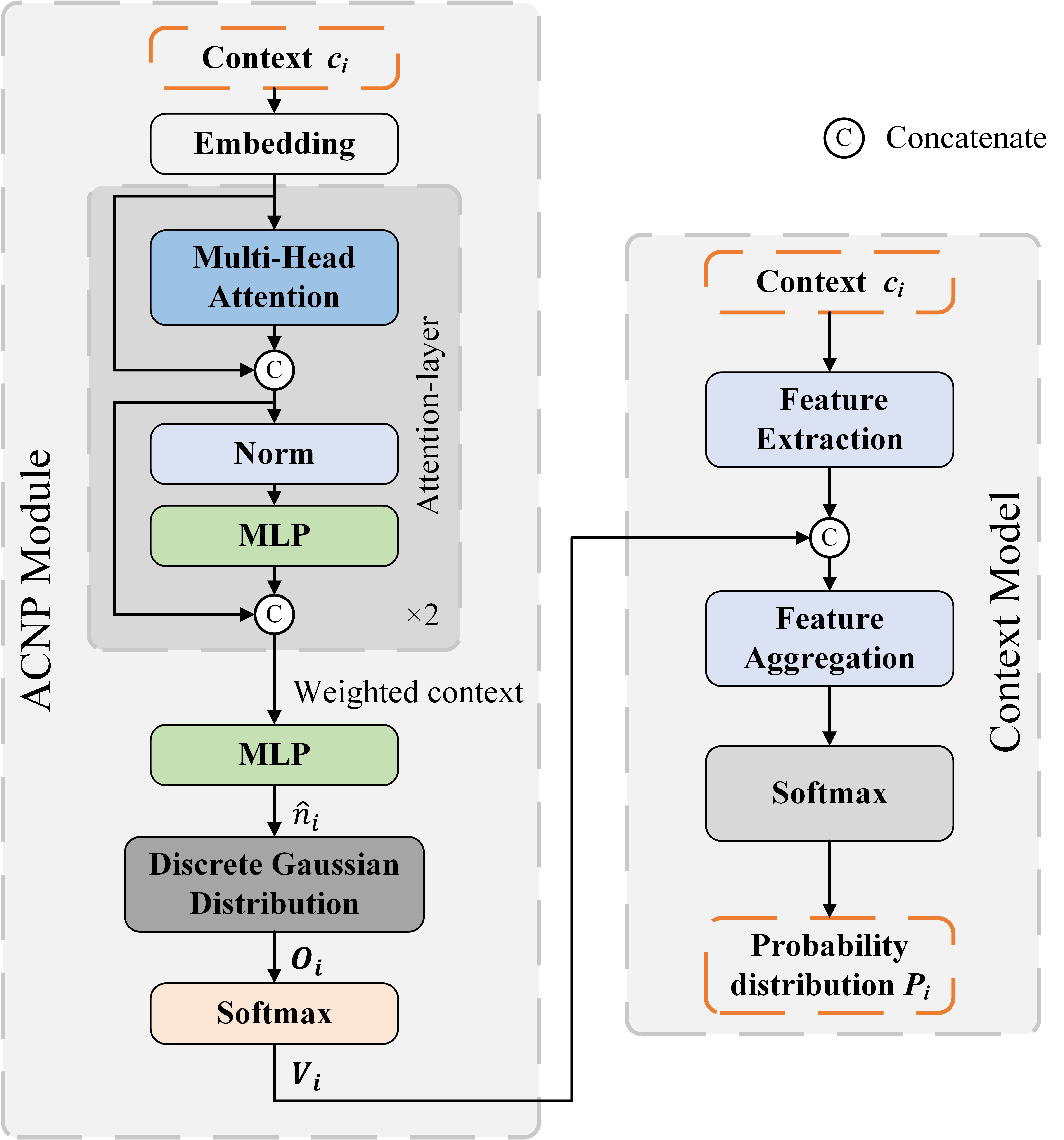}
\vspace{-0.2cm}%%减小图片上间隔
\caption{Overall architecture of the ACNP module. The ACNP module consists of attention layers and MLP layers. The ACNP module takes the context as input and outputs a vector containing information about the number of occupied child nodes. We divide the context model into two stages: feature extraction and feature aggregation. The dimension of features gradually decreases in the feature aggregation stage. The vector output by the ACNP module is concatenated with the output of the feature extraction stage and fed into the feature aggregation stage. The output of the context model is a 255-dimensional probability distribution used for entropy coding.}
\vspace{-0.4cm}
\label{fig1}
\end{figure} 

\subsection{Attention-based Child Node Number Prediction Module}

To address the aforementioned problem, we propose a general module, called ACNP, based on the attention mechanism \cite{24}. ACNP aims to enhance octree-based context models by introducing the predicted number of occupied child nodes into the context models. The structure of ACNP is shown in Fig. 2. The context is first fed into an attention layer to produce a weighted context. Then, a two-layer MLP aggregates the weighted context and outputs the predicted number of occupied child nodes. To adapt to diverse context models, the context fed into the ACNP module is the same as the context fed into the context model. The predicted number $\hat{n}_i$ of occupied child nodes of $\boldsymbol{X_i}$ can be expressed as

\begin{equation}
\hat{n}_i = M(\boldsymbol{c_i};\boldsymbol{\theta}),
\end{equation}
where $M$ denotes the ACNP module, $\boldsymbol{c_i}$ is the context of node $\boldsymbol{X_i}$, and $\boldsymbol{\theta}$ is the set of parameters of the ACNP module. As a one-dimensional feature is not suitable for the training of MLP, we expand the dimension of $\hat{n}_i$ while preserving its information about the predicted number of occupied child nodes. Specifically, we map $\hat{n}_i$ to an 8-dimensional vector $\boldsymbol{O_i}$ as follows

\begin{equation}
O_i(k) = \frac{1}{\sqrt{2\pi\sigma_i^2}} \exp\left(-\frac{(k-\mu_i)^2}{2\sigma_i^2}\right)\quad,
\end{equation}
where $\sigma_i=1$, $\mu_i=\lceil \hat{n}_i \rceil$, and $k$ is an integer in $[1,8]$. We then feed $\boldsymbol{O_i}$ into a Softmax layer and obtain an 8-dimensional vector $\boldsymbol{V_i}$. The $k$-th dimension of $\boldsymbol{V_i}$ represents the probability that node $\boldsymbol{X_i}$ has $k$ occupied child nodes. Subsequently, $\boldsymbol{V_i}$ can be inserted into the context model, as shown in the right part of Fig. 2. To accommodate the varied network structures among different context models, $\boldsymbol{V_i}$ is introduced into the feature aggregation stage, where the dimension of features starts to decrease drastically. Finally, the probability distribution predicted by the context model of node $\boldsymbol{X_i}$ can be expressed as
\begin{equation}
\boldsymbol{P_i}=G(\boldsymbol{c_i},\boldsymbol{V_i};\boldsymbol{\omega}).
\vspace{-0.4cm}
\end{equation}

\subsection{Loss Function}
As predicting the number of occupied child nodes is a regression problem, the ACNP module is trained using the Mean Squared Error (MSE) loss:

\begin{equation}
L_{MSE} = \frac{1}{m} \sum_{i=1}^{m} (n_i - \hat{n}_i)^2,
\end{equation}
where $n_i$ is the actual number of occupied child nodes.
The context model is still trained using the cross-entropy loss:
\begin{equation}
L_{CE} = -\sum_{i=1}^{m}\log_2{p_i}.
\end{equation}

\section{EXPERIMENTS AND DISCUSSION}
To verify the efficiency of the proposed ACNP module, we used it to enhance OctAttention \cite{21} and OctSqueeze \cite{18}. The resulting models are named as ACNP-OctAttention and ACNP-OctSqueeze, respectively. To validate their performance, we compared ACNP-OctAttention with OctAttention and EM-OctAttention \cite{23}, and compared ACNP-OctSqueeze with OctSqueeze on the Microsoft Voxelized Upper Body (MVUB) \cite{25}, 8i Voxelized Full Bodies (MPEG 8i) \cite{26}, and SemanticKITTI\cite{27} point cloud datasets. The point clouds in the first two datasets represent humans and are dense, while the point clouds in the third dataset were captured with LiDAR sensors and are sparse.

\subsection{Experimental Details}
In our first experiment, we trained OctAttention, OctSqueeze, and ACNP using \textit{andrew10, david10}, and {sara10} from MVUB, as well as \textit{soldier10} and \textit{longdress10} from MPEG 8i, and tested them on all remaining point clouds in these two datasets. All point clouds had 10-bit geometry precision, except for the test point clouds boxer12 and thaidancer12 from MPEG 8i, which had higher precision. As these two point clouds were downsampled in [21] to 10 and 9 bits, respectively, we did the same to ensure consistent comparison. In this experiment, the point clouds were compressed losslessly. Therefore, we used the bitrate to compare the performance of each method.

In our second experiment, we considered the SemanticKITTI dataset. The point clouds were first partitioned into 12-level octrees. Point cloud sequences 00 to 10 were used for training, and point cloud sequences 11 to 21 were used for testing. As partitioning point clouds into octrees introduces distortion, and the compression of the octree is lossless, point clouds reconstructed by different context models are identical. Therefore, the performance of each model is solely determined by the bitrate.    

We implemented ACNP-OctAttention and ACNP-OctSqueeze in PyTorch. The ACNP module and context models were trained separately. The ACNP module was trained for 20 epochs using an Adam optimizer with a learning rate of 0.001 and a decay of 0.93. All hyperparameter settings related to OctAttention were as in \cite{21}. Specifically, OctAttention was trained for 80 epochs using an Adam optimizer with a learning rate of 0.001 and a decay of 0.95. The context was composed of $N = 1024$ sibling nodes and $K = 4$ ancestor nodes. The 8-dimensional vector output from the ACNP module was concatenated with the output form the first layer of MLP in OctAttention.

As in \cite{18}, OctSqueeze was trained for 40 epochs using an Adam optimizer with a learning rate of $1e-4$. The context was composed of $K = 4$ ancestor nodes. The 8-dimensional vector was concatenated with the output from the second MLP in OctSqueeze.

Both training and testing were conducted on an Intel(R) Xeon(R) Gold 6148 CPU and a GeForce RTX 4090 GPU with 24 GB memory.

\begin{table*}
\begin{center}
  \caption{Average bitrate in bits per input point (BPIP)}
  \label{tab1}
  \begin{tabular}{cc|cccc}
    \hline\rule{0pt}{8pt}
    Dataset&	Point cloud&ACNP-OctAttention&\multicolumn{3}{c}{Gain over} \\
    &&BPIP	&G-PCC &	OctAttention &	EM-OctAttention \\
    \hline

\multirow{7}*{8i}&Loot&0.596&-19.24\% &-1.16\%&-0.82\%\\
~ &RedandBlack &0.706&-15.04\%&-1.55\%&-1.33\%\\
 & Thaidancer10&	0.623&-12.01\%& -1.27\%&-1.25\%\\
~&Thaidancer9 &	0.617&-15.13\%&-2.53\%&-2.53\%\\
~ &Boxer10	&0.564&-14.29\%& -0.70\%&-0.05\%\\
~&Boxer9&	0.575 &	-19.27\% &-0.17\% &-0.12\%\\
~&Average&	0.613 &	-17.07\%&-1.35\% &-1.07\%\\
\hline
\multirow{5}*{MVUB} &Phil10&0.761 &	-19.89\%&-2.65\%&-1.49\%\\
~&Phil9	&0.804 &-19.52\%&-2.11\% &-1.62\%\\
~&Ricardo10	&0.688 	&-22.70\% &-1.88\%	&-1.43\% \\
~&Ricardo9	&0.689	&24.29\%&-2.20\%&-1.75\%\\
~&Average	&0.736 	&-21.60\%&-2.21\%&-1.57\%\\
\hline
\multirow{1}*{SemanticKITTI} &Sequences 11-21&3.50 &-20.9\%&-3.05\%&-0.85\%\\
    \hline
  \end{tabular}
  \vspace{-0.5cm}
\end{center}
\end{table*}

\begin{table*}
\begin{center}
  \caption{Average bitrate in bits per input point (BPIP)}
  \label{tab1}
  \begin{tabular}{cc|cc}
    \hline\rule{0pt}{8pt}
    Dataset&	Point cloud &\multicolumn{1}{c}{ACNP-OctSqueeze}&Gain over\\
    &&BPIP   & OctSqueeze	\\
    \hline

\multirow{7}*{8i}&Loot&0.830&-1.07\%\\
~ &RedandBlack &0.947 &-1.35\%\\
 & Thaidancer10&0.859 & -0.69\%\\
~&Thaidancer9 &0.875 & -0.79\%\\
~ &Boxer10	& 0.805 & -1.95\%\\
~&Boxer9&0.827 & -0.24\%\\
~&Average&0.854 &	 -1.21\%\\
\hline
\multirow{5}*{MVUB}&Phil10&1.04 &-1.60\%\\
~&Phil9	&1.06 &-0.84\%\\
~&Ricardo10	& 0.963 &-1.23\% \\
~&Ricardo9	&0.937 &-1.47\%\\
~&Average	&1.00 &-1.28\%\\
\hline
\multirow{1}*{SemanticKITTI} &Sequences 11-21&4.43&-2.32\%\\
    \hline
  \end{tabular}
  \vspace{-0.5cm}
\end{center}
\end{table*}

\begin{table*}
\begin{center}
  \caption{Computational complexity comparison}
  \label{tab7}
  \begin{tabular}{c|ccccc}
    \hline\rule{0pt}{8pt}
    &OctAttention& EM-OctAttention	&Increase& ACNP-OctAttention &	 Increase\\
    \hline
  Model size (M)	  &28.0	 &29.3 &4.6\%  &55.4 &97.8\% \\
Encoding time (s)	&0.263	&0.274	&4.2\%&0.292&11.0\%\\
Decoding time (s)	&345	&362	&4.9\% &389 &12.8\%\\

    \hline
  \end{tabular}
  \vspace{-0.5cm}
\end{center}
\end{table*}

\begin{table*}
\begin{center}
  \caption{Computational complexity comparison}
  \label{tab7}
  \begin{tabular}{c|ccc}
    \hline\rule{0pt}{8pt}
    &	OctSqueeze&	ACNP-OctSqueeze&Increase\\
    \hline
  Model size (M)	  &26.7	 &52.8 &97.7\%\\
Encoding time (s)   &1.93 &2.11 &9.3\%\\
Decoding time (s)	&3.78 &4.39&16.1\%\\

    \hline
  \end{tabular}
  \vspace{-0.5cm}
\end{center}
\end{table*}

\subsection{Experimental results}
As shown in Table \uppercase\expandafter{\romannumeral1}, ACNP-OctAttention yielded the lowest bitrate on MVUB, MPEG 8i, and SemanticKITTI. ACNP-OctAttention outperformed the G-PCC test model TMC13 v23.0 \cite{28} by reducing the bitrate by 17.07\% on MPEG 8i and 21.6\% on MVUB. It also reduced the bitrate by 1.35\% on MPEG 8i and 2.21\% on MVUB compared to OctAttention. These results indicate that introducing the ACNP module enhances the performance of the context model. Furthermore, ACNP-OctAttention outperformed EM-OctAttention by reducing the bitrate by 1.07\% on MPEG 8i and 1.57\% on MVUB.On SemanticKITTI, ACNP-OctAttention reduced the bitrate by 3.05\% compared to OctAttention and 0.85\% compared to EM-OctAttention. These results show that including the predicted number of occupied nodes leads to better performance than including the predicted occupancy. In addition, as shown in Table \uppercase\expandafter{\romannumeral2} ACNP-OctSqueeze also reduced the bitrate by 1.28\% on MPEG 8i, 1.21\% on MVUB, and 2.32\% on SemanticKITTI compared to OctSqueeze. These experimental results show that the ACNP module is general and can enhance the performance of existing context models.

\subsection{Complexity Comparison}
Table \uppercase\expandafter{\romannumeral3} and Table \uppercase\expandafter{\romannumeral4} presents the sizes of the models and their average coding times on the MPEG 8i dataset. Due to the inclusion of the ACNP module, ACNP-OctAttention was 97.8\% larger than OctAttention, resulting in a 11.0\% increase in encoding time and a 12.8\% increase in decoding time. Similarly, ACNP-OctSqueeze was 97.7\% larger than OctSqueeze, leading to a 9.3\% increase in encoding time and a 16.1\% increase in decoding time. In our implementation, to achieve efficiency, computations in the ACNP module are executed concurrently with feature extraction in the context model.

\section{Conclusion}

We first showed that the cross-entropy loss is not suitable for measuring the difference between the predicted probability distribution and the label when the number of occupied child nodes is unknown. Subsequently, we proposed an attention-based module (ACNP) to predict the number of occupied child nodes, thus assisting in the prediction of the probability distribution within the context model. Experimental results showed that ACNP significantly improved the coding efficiency of two octree-based context models: OctAttention and OctSqueeze. However, ACNP increased the complexity of the context models. Future work may include reducing the complexity of ACNP and applying it to other context models.

\end{document}